# Large planets may not form fractionally large moons

Miki Nakajima [1,2 ✉], Hidenori Genda [3], Erik Asphaug [4] & Shigeru Ida [3]

One of the unique aspects of Earth is that it has a fractionally large Moon, which is thought to have formed from a Moon-forming disk generated by a giant impact. The Moon stabilizes the Earth's spin axis at least by several degrees and contributes to Earth's stable climate. Given that impacts are common during planet formation, exomoons, which are moons around planets in extrasolar systems, should be common as well, but no exomoon has been confirmed. Here we propose that an initially vapor-rich moon-forming disk is not capable of forming a moon that is large with respect to the size of the planet because growing moonlets, which are building blocks of a moon, experience strong gas drag and quickly fall toward the planet. Our impact simulations show that terrestrial and icy planets that are larger than ~1.3—1.6$R_⊕$ produce entirely vapor disks, which fail to form a fractionally large moon. This indicates that (1) our model supports the Moon-formation models that produce vapor-poor disks and (2) rocky and icy exoplanets whose radii are smaller than ~1.6$R_⊕$ are ideal candidates for hosting fractionally large exomoons.

[1] Department of Earth and Environmental Sciences, University of Rochester, P.O. Box 270221, Rochester, NY 14627, USA. [2] Department of Physics and Astronomy, University of Rochester, P.O. Box 270171, Rochester, NY 14627, USA. [3] Earth-Life Science Institute, Tokyo Institute of Technology, 2-12-1, Ookayama, Meguro-ku, Tokyo 152-8550, Japan. [4] University of Arizona, Lunar and Planetary Laboratory, 1629 E. University Blvd, Tucson, AZ 85721, USA. ✉email: mnakajima@rochester.edu

Earth is the only known life-hosting planet and has a number of features unique to our solar system, including active plate tectonics, a strong magnetic field, and a large moon with respect to the size of Earth. The presence of the Moon controls the length of the day and ocean tides, which affects the terrestrial biological cycles. The Moon also stabilizes the Earth's spin axis by at least several degrees[1–3]. Thus, at least for Earth, the Moon also contributes to Earth's stable climate and potentially offers an ideal environment for life to develop and evolve. It should be noted, however, that the stability depends on several factors, such as the initial obliquity of the planet[4,5].

The origin of the Moon and conditions to host a massive satellite have been under active debate. According to the canonical hypothesis of the lunar origin, the Moon formed from a partially vaporized disk generated by a collision between Earth and a Mars-sized impactor approximately 4.5 billion years ago[6–8]. This giant impact hypothesis was originally thought to be able to explain the geochemical observation that the Earth and Moon have nearly identical isotopic ratios[9], assuming that a large impact can homogenize the two bodies. However, numerical simulations find that the disk is primarily made of the impactor materials and it is challenging to mix the two reservoirs[10]. This requires the impactor ("Theia") to have isotopic ratios nearly identical to Earth by chance[11,12] or homogenization during the disk stage[13].

Alternatively, the Moon could have formed by a more energetic impact that would have homogenized the system more easily. Proposed energetic impacts include a collision between two half-Earth-sized objects[14], and other high-energy and high-angular momentum cases[15,16]. One of the potential problems of the energetic models is that the final angular momentum of the system is two to three times higher than the current value, which may not be easily removed from the system[17]. Moreover, these impacts would have homogenized the Earth's mantle, which may not be consistent with geochemical observations[18]. Another option is that the Moon could have formed by multiple impacts[19]. Small- and high-speed impactors can strip and launch Earth's mantle materials into orbit, but if some of the impactors are large or slow, the disk and therefore the Moon would still hold isotopic signatures originating from the impactors.

These impacts generate the Moon-forming disks with various vapor mass fractions (VMF). In general, energetic impacts lead to higher VMF of the disk. The VMF for the canonical model is 0.1–0.3[20], and those for half-Earths as well as high velocity and high-angular momentum cases are 0.7–1.0[20], and those for the multiple moon model are 0.1–0.5[19]. As the disk cools, VMF decreases. It is conventionally thought that moonlets (or lunatesimals) accrete from the liquid portion of the disk and moonlets accrete into the Moon over time. We revisit this process later in this work.

Given that planetary collisions are common in the young solar system and may have already been detected in extrasolar systems (e.g., refs. [21–24]), it is expected that impact-induced moons (exomoons) are common (e.g., ref. [25]). Unfortunately, we currently have a very limited understanding of the origin of exomoons due to a lack of confirmed exomoons. While several detection methods have been proposed for exomoons, including direct imaging[26] and microlensing[27], the Hunt for Exomoons with Kepler (HEK) (e.g., ref. [28]) is leading the effort for exomoon detection, analyzing the Kepler data for an indication of any transit timing and transit duration variations (TTV and TDV) of exoplanets due to exomoons. Despite the extensive search (57 exoplanets were searched during HEK I-V[28]) and a follow-up study with the Hubble Space Telescope[29], no exomoon has been confirmed to date[30–32]. Detecting an exomoon is very challenging, and the lifetime of moons around close-in planets, such as most of the Kepler planets, could be short due to the strong tide from the stars[33]. Moreover, close-in planets have small Hill radii, which is the distance where the gravity of the planet is dominant, and some moons that migrate outward can be lost due to orbital instability[33]. In the canonical "graze and merge" moon-forming scenario[8], the return of the projectile occurs only if the projectile loops inside the 1/3 of the Hill radius[34]. Hence closer-in planets (with smaller Hill radii) would be less likely to experience a similar graze-and-merge impact. Nonetheless, planetary impacts that would lead to moon formation should be common in the late stages of planetary accretion, and the lack of a confirmed exomoon detection has not been fully explained.

Here, we explore the possibility of not all the planets can form impact-induced moons, which can be fractionally large with respect to their host planets. We constrain the condition to form impact-induced moons in order to (1) constrain the lunar origin, and (2) explain the lack of confirmed exomoons to date. We conduct impact simulations with the smoothed particle hydrodynamics (SPH) method, which describes fluid as a collection of particles[20,35], to characterize moon-forming disks and moon accretion process. We consider two endmembers of planetary compositions: a rocky planet consisting of 70 wt% of forsterite mantle and 30 wt% iron core and an icy planet consisting of 70 wt% water ice and 30 wt% forsterite core. Gas giants and mini-Neptunes are not the primary focus of this study, but we discuss implications for these planets later. Our simulations show that a moon-forming disk with a high vapor mass fraction (VMF) is not capable of forming a moon that is large with respect to the size of the planet because growing moonlets of 100 m–100 km in size in the disk experience strong gas drag from vapor, lose their angular momentum and fall onto the planet on a short timescale, failing to grow further. This does not occur in a vapor-poor disk (i.e., small VMF) because the gas drag effect is much weaker. An initially vapor-rich disk can start forming stable moonlets once it cools and the VMF is small enough, but by the time a significant portion of the disk mass is lost, failing to form a fractionally large moon. Consequently, our study supports the moon-formation models that produce initially small VMF, such as the canonical model. Moreover, we find that rocky planets larger than $6\,M_\oplus$ (~$1.6\,R_\oplus$, where $R_\oplus$ is the Earth radii), where $M_\oplus$ is the Earth mass, and icy planets larger than $1\,M_\oplus$ (~$1.3\,R_\oplus$) produce completely vapor disks, and therefore these planets are not capable of forming fractionally large impact-induced moons. For this reason, we propose that future exomoon observations should focus on exoplanets smaller than ~$1.6\,R_\oplus$ to detect impact-induced exomoons.

## Results

**Effect of vapor on growing moonlets.** If a moon-forming disk started from a completely vapor state (i.e., VMF = 1.0), liquid droplets would start forming as the disk cools. These liquid droplets fall toward the midplane on a short timescale[36]. They would experience strong gas drag from the vapor because droplets orbit around the planet with the Keplerian velocity $v_K$, whereas vapor moves slower due to the radial pressure support. This radial drift velocity is described as $v_r = -2\eta v_K \tau_f/(1+\tau_f^2)$[37], where $\eta = -\frac{1}{2}(\frac{c_s}{v_K})^2 \frac{\partial \ln P}{\partial \ln r}$ is the pressure gradient parameter, which describes the strength of the radial pressure gradient of the vapor, $P$ is the pressure, $c_s$ is the sound velocity, $v_K$ is the Keplerian velocity, $\tau_f$ is the dimensionless stopping time expressed as $\tau_f = \frac{8}{3C_D\eta}\frac{\rho_p}{\rho_g}\frac{R_p}{r}$, where $C_D$ is the drag coefficient, $\rho_p$ is the particle density, $\rho_g$ is the gas density, $R_p$ is the particle radius, and $r$ is the distance from the planet. The radial fall velocity is largest when $\tau_f = 1$ and potentially the particle can fall toward the planet very quickly, depending on the disk condition. This gas drag effect has been considered for small droplets, but not for growing moonlets[16], which are building blocks of a moon.

In fact, this was a major challenge for planet formation in the protoplanetary disk. The very same mechanism would have removed 1 m-sized planetesimals, which are building blocks of planets, at 1 AU on the timescale of 80 years[37,38], which is a much shorter timescale than the planet formation timescale (a few to tens of millions of years). This is the so-called "meter-barrier problem" and was an outstanding problem in planetary science for decades.

While this is the first time to investigate the effect of gas drag in a moon-forming disk in detail, previous work also points out another issue of a vapor-rich disk. The previous work[39] suggests that an initially vapor-rich moon-forming disk may not be dynamically stable due to shocks caused by density gradients in the disk and loses a significant portion of its mass in a short time, concluding that a vapor-rich moon-forming disk is not capable of forming a large moon.

**SPH simulations.** We conduct SPH simulations to quantify VMF as well as the pressure graduate parameter $\eta$ to characterize the strength of the gas drag. In the case of rocky planets, including Earth, a high VMF of a moon-forming disk means that the disk is primarily made of silicate vapor and a small amount of liquid (magma). For an icy planet, a high VMF can mean a steam disk potentially with embedded silicate droplets. The parameters for SPH simulations are the total (target and impactor) mass $M_T$ ($M_\oplus \leq M_T \leq 6 M_\oplus$ for rocky planets and $0.1 M_\oplus \leq M_T \leq 1 M_\oplus$ for icy planets, where $M_\oplus$ is the Earth mass), the impactor-to-total-mass ratio $\gamma$ ($0.13 \leq \gamma \leq 0.45$), impact velocity $v_{imp}$, and impact angle $\theta$. In this work, we use the fixed impact angle ($\theta = 48.6°$) and the impact velocity ($v_{imp} = v_{esc}$), where $v_{esc}$ is the mutual escape velocity. The impact angle and velocity are similar to those for the canonical moon-forming impact models[8,10]. The reason why we explore different mass ranges for the rocky and icy planets is that the required mass for complete vaporization is different between them, as we discuss in detail below.

Figure 1 shows snapshots of giant impacts with SPH. The top two rows represent an impact between rocky planets and the bottom two rows represent an impact between icy planets. The red-yellow and blue-sky-blue colors represent mantle entropy. Gray (iron) and orange (silicate) represent the core materials. In both simulations, the same input parameters are used ($M_T = 1 M_\oplus$, $\gamma = 0.13$, $\theta = 48.5°$, $v_{imp} = v_{esc}$). The icy planets are larger than rocky planets because ice is much less dense than silicate. The overall dynamics during an impact are remarkably similar between the two systems (Fig. 1). In contrast, thermodynamics during the impacts differ significantly. The overall disk vapor mass fraction is approximately 0.3 (i.e., 70 wt% of the mass is in the liquid phase) for the rocky planet case, whereas it is 1.0 for the icy planet case. This is primarily because the latent heat of water (~$2.3 \times 10^6$ J kg$^{-1}$) is much smaller than the latent heat of silicate (~$1.2 \times 10^7$ J kg$^{-1}$ [40]).

The SPH output is summarized in Table 1. Figure 2 shows the disk mass $M_D$ and disk angular momentum $L_D$ as a function of the total mass $M_T$ for rocky and icy planets. The colors represent $\gamma$, and the dashed lines represent analytical estimates of the disk mass[25]. Generally, the disk mass agrees with the analytical model, especially at small $\gamma$. There is a slight upward trend as a function of the total mass in the disk mass and angular momentum at $\gamma \geq 0.30$. This may be because the larger extent of tidal deformation of the colliding bodies at the time of the impact with larger $M_T$, which may make the impact angle slightly shallower and generate a larger disk mass and a larger angular momentum. However, the deviation from the analytical solution is relatively small and more simulations with larger $M_T$ are needed to confirm this trend.

Figure 3 shows the VMF of a moon-forming disk as a function of $\gamma$ and $M_T$. The panels (a) and (b) represent rocky and icy planetary impacts. The vapor mass fraction is over 0.96 at the planetary mass $\geq 6 M_\oplus$ for rocky bodies and $\geq 1 M_\oplus$ for icy planets. The larger $M_T$ is, the higher VMF is in general. This is because the kinetic energy involved in such an impact generally increases as the planetary mass increases (see "Methods"). Moreover, a fractionally larger impactor (i.e., larger $\gamma$) contributes to higher impact-induced heating and higher VMF as well. An icy planet produces a disk with a higher VMF than a rocky planet with the same mass, as discussed above.

The disk structures at the midplane formed by the rocky and icy planet collisions are shown in Fig. 4 (Run ID6, $M_T = 5 M_\oplus$, $\gamma = 0.13$) and Fig. 5 (Run ID20, $M_T = 1 M_\oplus$, $\gamma = 0.13$), respectively. The dotted line in panel (a) represents the surface density directly obtained from the SPH simulation. We find that this initial disk structure is unstable because it does not meet the Rayleigh criterion where the disk angular momentum $L_z$ needs to monotonically increase in the radial direction ($dL_z/dr > 0$, where r is the radial distance from the planet)[20,41]. Thus, the disk would dynamically adjust itself within a few days. Two potential disk surface density structures shown in the solid lines satisfy this condition while conserving the disk angular momentum and mass. The skye-blue line represents an exponential model and the green line represents a model that is a polynomial model (see "Methods"). The disk temperature (panel b), vapor mass fraction (panel c) and the pressure gradient parameter $\eta$ (panel d) are shown as a function of $r/R_\oplus$, where $R_\oplus$ is the Earth radius. The temperature, vapor mass fraction, and $\eta$ are calculated assuming that the disk is in the hydrostatic equilibrium[20]. The pressure parameter $\eta$ in the disk is typically ~0.02–0.06 for rock and icy moon-forming disks, which is ~10 times higher than the typical values in the protoplanetary disk[42].

**Radial fall of moonlets.** Assuming that $\eta \sim 0.04$, the radial fall timescale for this disk is ~1 day for a 2-km-sized body for a silicate vapor disk and ~1 day for a 1.3 km for a water vapor disk (see Methods). This is comparable to the vertical fall timescale of particles (~a few days[36]) and much shorter than the accretion timescale of solid and liquid materials (~1–10s years[43]). The typical lifetime for the moon-forming disk is estimated to be ~100 years[44], and therefore the radial fall timescale is extremely short for a few km-sized moonlets. This process can prevent moonlets from growing larger than km in size, which we call here the "km-barrier problem" for moon-forming disks. One of the potential differences between the meter-barrier and km-barrier problems is that mutual gravity can facilitate the growth of the moonlets in the latter scenario, but this effect is likely minor and would not solve the radial fall issue.

An initially 100 wt% vapor disk cannot form a moon that is large with respect to the size of the planet due to the rapid infall timescale, according to our arguments above. We, therefore, propose that the moon-forming disk needs to be initially vapor-poor, supporting the canonical moon-formation hypothesis whose vapor mass fraction is small (0.1–0.3). Moreover, we can constrain the large moon-forming region as shown in Fig. 6; rocky planets larger than 6 $M_\oplus$ and icy planets larger than 1 $M_\oplus$ may not form a large satellite because their VMFs are 1.0. Here, a "large" moon is defined as a satellite whose mass is approximately a few to 10 wt% of the planetary mass. This is consistent with the Earth–Moon (VMF ~0.1–0.3[20]), and Pluto–Charon (icy planets, VMF ~ 0–0.08[45]), given that these moon-forming impacts produce small VMFs. The radii of a 6 $M_\oplus$ rocky planet and a 1 $M_\oplus$ icy planet are ~1.3–1.6 $R_\oplus$[46,47]. We also predict that gaseous mini-Neptunes may also have trouble producing large moons because of their large masses as well as potentially more vapor-rich disks if the H/He gas becomes part of the disk[48].

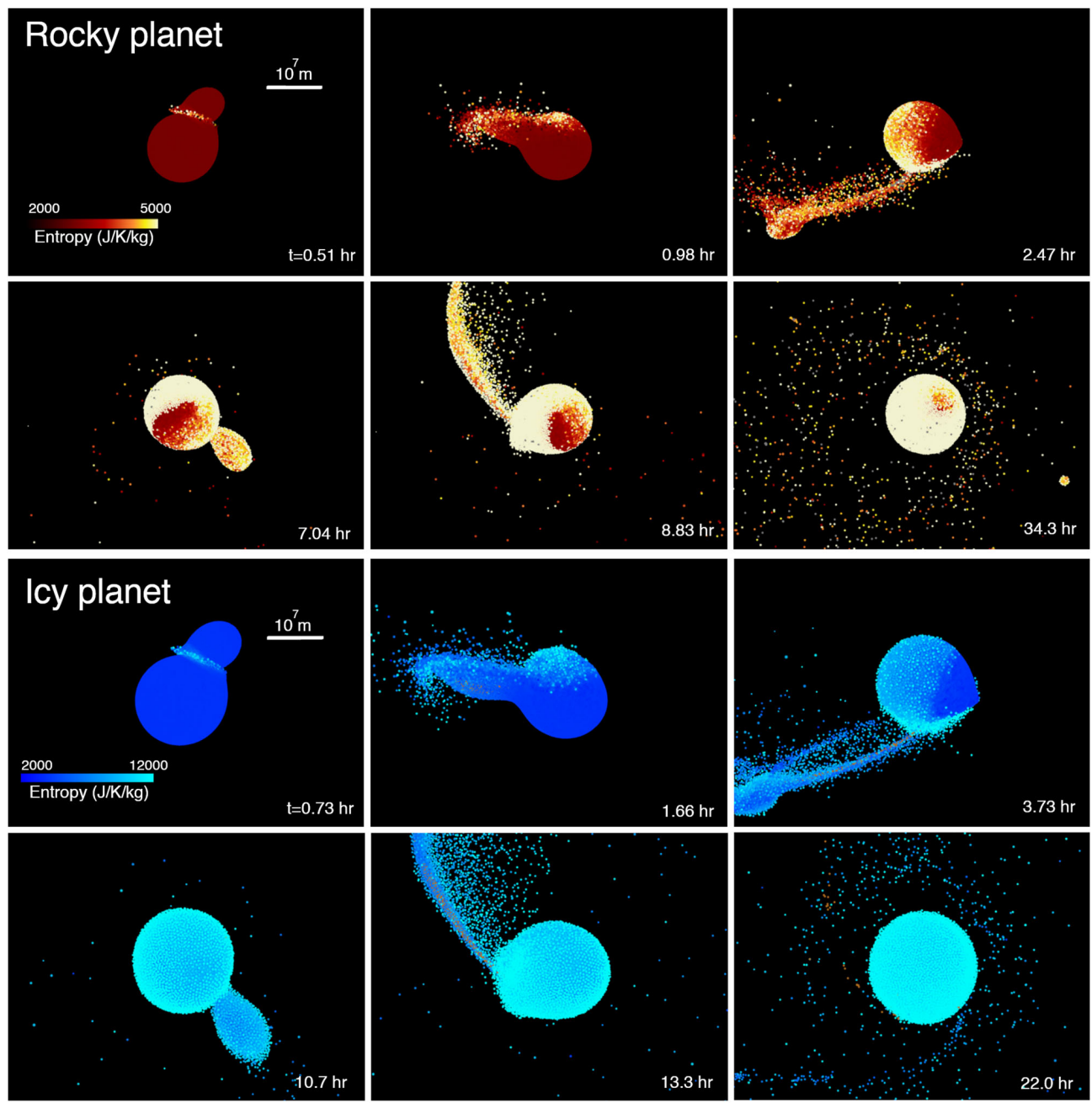

**Fig. 1 Snapshots of giant impacts (Runs ID2 and ID20, Table 1).** The top two rows represent an impact between two rocky planets. The red-orange colors represent the entropy of the mantle material (forsterite). The iron core is shown in gray. The bottom two rows represent an impact between two icy planets. The blue-sky-blue colors represent the entropy of water ice, and the orange color represents forsterite. The scale represents $10^7$ m.

### Discussion
It should be noted that our model does not exclude forming small moons around large planets. As the disk cools and the VMF decreases, the gas drag effect weakens and moonlets start to form. However, during the cooling process, the vapor-rich disk would viscously expand and a significant portion of the disk would be lost to the planet (see "Methods" for detailed discussion). This can be an explanation for fractionally small moons of Uranus ($1.02 \times 10^{-4} M_p$)[49], which could have formed by a large impact[48–50].

A large impact is not the only way to form a moon. A satellite can also form within a circumplanetary disk. The Galilean satellites and Titan are thought to have formed from dust in such circumplanetary disks. These disks typically have smaller dust-to-gas ratios (~$10^{-2}$–$10^{-4}$ (see ref. [51])). The typical satellite total mass formed in a circumplanetary disk is estimated to be $10^{-4} M_p$[51] due to the two competing processes: supply of dust materials to the disk and satellite loss due to their radial fall caused by the gas. This total mass estimate is also consistent with the newly observed exomoon disk mass (~$10^{-4} M_p$)[52]. A moon can form by gravitational capture. It is unclear if there is a maximum satellite size in the case of gravitational capture, but at least in the solar system, gravitationally captured moons have fractionally small masses (e.g., Triton, $2.46 \times 10^{-4} M_p$)[53]. A

Table 1 Summary of the results.

| ID | EOS | $M_T/M_\oplus$ | $\gamma$ | $v_{esc}$ (km s$^{-1}$) | $M_p/M_T$ | $M_D/M_t$ | $L_D/L_T$ | $S_{ave}$ (J K$^{-1}$ kg$^{-1}$) | VMF | N |
|---|---|---|---|---|---|---|---|---|---|---|
| 1 | R | 1.0 | 0.13 | 9.282 | 0.9952 | 0.014 | 0.145 | 5231.9 | 0.348 | 50,000 |
| 2 | R | 1.0 | 0.13 | 9.282 | 0.9976 | 0.014 | 0.161 | 5341.8 | 0.372 | 100,000 |
| 3 | R | 2.0 | 0.13 | 11.745 | 1.955 | 0.011 | 0.127 | 5774.6 | 0.49 | 50,000 |
| 4 | R | 2.9 | 0.13 | 13.488 | 2.845 | 0.015 | 0.166 | 6710.8 | 0.801 | 50,000 |
| 5 | R | 4.0 | 0.13 | 15.041 | 3.907 | 0.014 | 0.169 | 6572.7 | 0.718 | 50,000 |
| 6 | R | 5.0 | 0.13 | 16.358 | 4.889 | 0.016 | 0.186 | 7103.7 | 0.885 | 50,000 |
| 7 | R | 6.0 | 0.13 | 17.490 | 5.849 | 0.016 | 0.181 | 7526.9 | 0.969 | 50,000 |
| 8 | R | 1.0 | 0.3 | 8.860 | 0.9538 | 0.057 | 0.248 | 5634.0 | 0.485 | 50,000 |
| 9 | R | 2.0 | 0.3 | 11.356 | 1.899 | 0.065 | 0.288 | 6279.5 | 0.705 | 50,000 |
| 10 | R | 3.0 | 0.3 | 13.156 | 2.835 | 0.071 | 0.311 | 6642.0 | 0.847 | 50,000 |
| 11 | R | 4.0 | 0.3 | 14.595 | 3.754 | 0.08 | 0.353 | 7098.7 | 0.955 | 50,000 |
| 12 | R | 5.0 | 0.3 | 15.822 | 4.682 | 0.084 | 0.365 | 7253.2 | 0.978 | 50,000 |
| 13 | R | 6.0 | 0.3 | 16.888 | 5.601 | 0.086 | 0.381 | 7545.0 | 0.996 | 50,000 |
| 14 | R | 1.0 | 0.45 | 8.790 | 0.9374 | 0.092 | 0.251 | 5871.5 | 0.587 | 50,000 |
| 15 | R | 2.0 | 0.45 | 11.257 | 1.868 | 0.11 | 0.311 | 6544.5 | 0.841 | 50,000 |
| 16 | R | 3.0 | 0.45 | 13.035 | 2.812 | 0.105 | 0.292 | 6658.9 | 0.888 | 50,000 |
| 17 | R | 4.0 | 0.45 | 14.465 | 3.728 | 0.115 | 0.31 | 6693.1 | 0.906 | 50,000 |
| 18 | R | 5.0 | 0.45 | 15.684 | 4.624 | 0.126 | 0.336 | 7116.3 | 0.986 | 50,000 |
| 19 | R | 6.0 | 0.45 | 16.764 | 5.541 | 0.129 | 0.346 | 7329.1 | 0.997 | 50,000 |
| 20 | I | 1.0 | 0.13 | 7.893 | 0.9675 | 0.02 | 0.206 | 11,842.2 | 1.0 | 100,000 |
| 21 | I | 1.0 | 0.13 | 7.893 | 0.9597 | 0.025 | 0.237 | 10,832.3 | 0.999 | 50,000 |
| 22 | I | 0.1 | 0.13 | 3.523 | 0.09683 | 0.027 | 0.289 | 6549.8 | 0.233 | 50,000 |
| 23 | I | 0.5 | 0.13 | 6.187 | 0.4883 | 0.018 | 0.2 | 10,022.1 | 0.926 | 50,000 |
| 24 | I | 0.1 | 0.3 | 3.406 | 0.09421 | 0.075 | 0.333 | 7286.2 | 0.35 | 50,000 |
| 25 | I | 0.5 | 0.3 | 6.003 | 0.4764 | 0.061 | 0.269 | 9387.0 | 0.876 | 50,000 |
| 26 | I | 1.0 | 0.3 | 7.6893 | 0.942 | 0.076 | 0.326 | 9785.4 | 0.955 | 50,000 |
| 27 | I | 0.1 | 0.45 | 3.386 | 0.09341 | 0.092 | 0.249 | 7583.8 | 0.411 | 50,000 |
| 28 | I | 0.5 | 0.45 | 5.942 | 0.4664 | 0.11 | 0.316 | 10,005.8 | 0.965 | 50,000 |
| 29 | I | 1.0 | 0.45 | 7.614 | 0.9425 | 0.095 | 0.284 | 11,244.5 | 0.998 | 50,000 |

ID is the simulation ID number, EOS represents the mantle element, where R represents a rocky planet and I represents an icy planet. $M_T$ is the total mass, $M_\oplus$ is the Earth mass, $\gamma$ is the impactor-to-total mass ratio, $v_{esc}$ is the escape velocity, $M_p$ is the planetary mass after the impact, $M_D$ is the disk mass, Mt is the target mass, $L_D$ is the disk angular momentum, $L_T$ is the total angular momentum of the system, $S_{ave}$ is the averaged disk entropy, VMF is the vapor mass fraction of the disk, and N is the number of SPH particles.

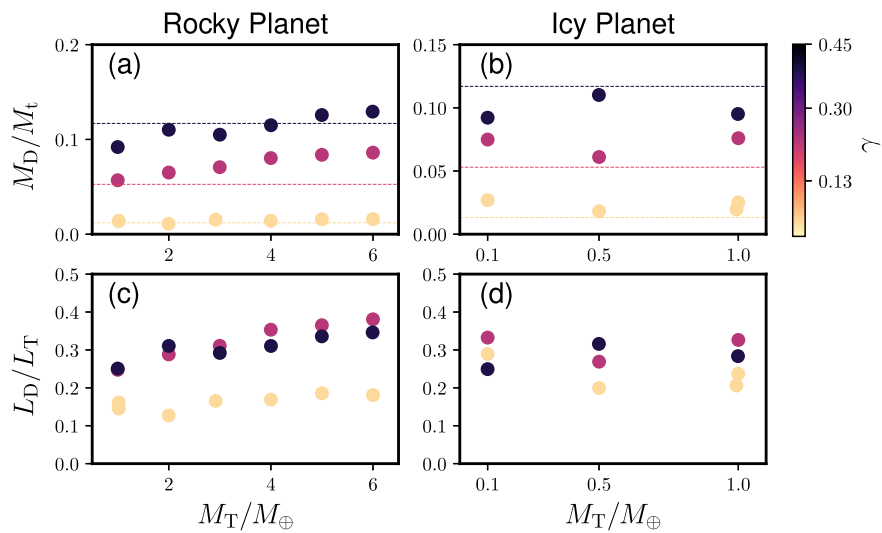

**Fig. 2 Disk masses and angular momenta based on our SPH simulations. a, b** The disk mass $M_D$ normalized by the target mass $M_t$ as a function of the total mass $M_T$ normalized by the Earth mass for rocky and icy planets, respectively. The colors represent the values of $\gamma$. The dashed lines represent the disk mass estimate from previous work assuming a rocky composition[25]. **c, d** The disk angular momentum $L_D$ normalized by the total angular momentum for rocky and icy planets $L_T$, respectively.

significant fraction of the orbital kinetic energy of a heliocentric body needs to be dissipated during capture, which requires certain conditions[53]. In contrast, impacts, especially at relatively low velocity, can naturally produce fractionally large moons (the Pluto–Charon system has 0.118 $M_p$). Under specific conditions, a nearly intact impactor can also be captured as a moon[45,54], but this requires very grazing incidence and very low relative velocity and is thus likely less frequent than moons formed in impact-induced disks, given that formation of an impact-induced disk is common[55].

There are several aspects that need to be considered further. The presence of the Roche radius near the moon-forming region makes the accretion process complex. Moonlets that enter inside the Roche radius become small fragments due to the planetary tide and they become well coupled with vapor. The evolution of the fragments may be similar to the evolution of the vapor disk (see "Methods"), but more detailed disk modeling is necessary to assess the effect of the Roche radius. In addition, if moonlets can skip this dangerous km size problem by forming a large moonlet very quickly, even a vapor-rich disk can potentially form a large moon. A promising mechanism is so-called streaming instability, which is a clump-formation mechanism due to the spontaneous concentration of particles followed by gravitational collapse. This mechanism was originally proposed to solve the planet formation problem in the protoplanetary disk[56]. Whether the same mechanism works for a moon-forming disk will be considered in our future study[57].

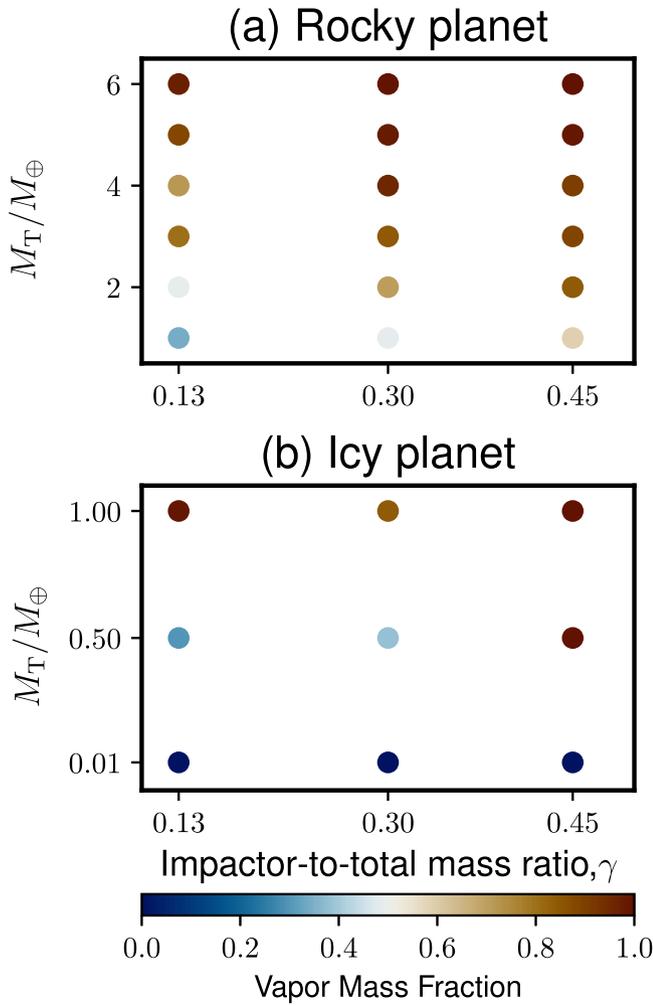

**Fig. 3 The vapor mass fraction of the disk as a function of the impactor-to-total mass ratio and the total mass normalized by the Earth mass.** The panels (**a**) and (**b**) represent rocky and icy planet cases, respectively. Brown colors represent higher vapor mass fractions, whereas blue colors represent lower values.

We only explore limited parameter space (e.g., impactor size and its velocity), but of course other parameters are possible. Nevertheless, the parameters chosen here tend to produce a disk that is massive enough to form a large satellite and that is less shock-heated than other scenarios (especially when $\gamma = 0.13$). Moreover, it is likely that high VMF cases (e.g., ≥0.8) are enough to inhibit the accretion of large moons, but the threshold VMF is difficult to determine. For this reason, we assume that VMF = 1 is the upper limit to form an impact-induced large moon.

Here, we consider collisions between planets that have the same compositions, but a collision between an icy and rock planets is possible. This may make it possible for a large icy planet (≥1 $M_\oplus$) to have a large rocky satellite if the disk is primarily made of the rocky impactor. Nevertheless, such impacts are likely less common unless there are extensive radial mixing events that lead to a large fraction of planetary objects crossing the ice line (e.g., Nice model and grand tack scenario[58,59]).

Our model predicts that the moon-forming disk needs to be initially liquid or solid rich, supporting the canonical moon-forming impact model[8]. Moreover, this work will help narrow down planetary candidates that may host exomoons; we predict that planets whose radii are smaller than ~1.6$R_\oplus$ would be good candidates to host fractionally large exomoons (≤6 $M_\oplus$ for rocky planets and ≤1 $M_\oplus$ for icy planets). These relatively small exoplanets are understudied (only four out of 57 exoplanets surveyed by the HEK project are under this radius limit), which can potentially explain the lack of exomoon detection to date. Super-Earths are likely better candidates than mini-Neptunes to host exomoons due to their generally lower masses and potentially lower H/He gas contribution to the disk. This narrower parameter space may help constrain exomoon search in data from various telescopes, including Kepler, the Hubble space telescope, CHaraterising ExOPlanet Satellite (CHEOPS)[60], and the James Webb Space Telescope (JWST).

## Methods

**Estimate on radial fall timescale.** In the moon-forming disk, the Newton drag force is written as $F = C_D \pi R_p^2 \rho_g v_{rel}^2 / 2$, where $C_D$ is the drag coefficient (~0.44 for turbulent flows), $R_p$ is the particle radius, $\rho_g$ is the gas density, $v_{rel}(= \eta v_K)$ is the relative velocity between the particles and gas[38], and $v_K$ is the Keplerian velocity. The friction time is expressed as $t_f = \frac{8}{3C_D} \frac{\rho_p R_p}{\rho_g v_{rel}}$ where $\rho_p$ is the particle density. The dimensionless stopping time $\tau_f = \Omega_K t_f$, where $\Omega_K$ is the Keplerian angular velocity (here $\Omega_K = \sqrt{GM_\oplus/(3R_\oplus)^3} = 2.38 \times 10^{-4}$ s$^{-1}$ at $r = 3R_\oplus$). The dimensionless stopping time is scaled as

$$\tau_f = \Omega_K t_f = \frac{8}{3C_D \eta} \frac{\rho_p}{\rho_g} \frac{R_p}{r}$$
$$= 1.18 \left(\frac{\eta}{0.04}\right)^{-1} \left(\frac{\rho_p}{3000 \text{ kg m}^{-3}}\right) \left(\frac{\rho_g}{40 \text{ kg m}^{-3}}\right)^{-1} \left(\frac{R_p}{2 \text{ km}}\right) \left(\frac{r}{3R_\oplus}\right)^{-1}. \quad (1)$$

The radial drift rate is described as $v_r = -2\eta v_K \tau_f/(1 + \tau_f^2)$[37]. The timescale for radial fall is

$$t_{\text{fall}} = \frac{r}{v_r} = \frac{1}{2\eta} \left(\frac{r^3}{GM}\right)^{\frac{1}{2}} \frac{1 + \tau_f^2}{\tau_f}$$
$$= \begin{cases} 12.3 \text{ h} \left(\frac{\rho_g}{40 \text{ kg m}^{-3}}\right) \left(\frac{\rho_p}{3000 \text{ kg m}^{-3}}\right)^{-1} \left(\frac{M}{M_\oplus}\right)^{-\frac{1}{2}} \left(\frac{r}{3R_\oplus}\right)^{\frac{5}{2}} \left(\frac{R_p}{2 \text{ km}}\right)^{-1} & \text{at } \tau_f \ll 1, \\ 29.1 \text{ h} \left(\frac{\eta}{0.04}\right)^{-1} \left(\frac{M}{M_\oplus}\right)^{-\frac{1}{2}} \left(\frac{r}{3R_\oplus}\right)^{\frac{3}{2}} & \text{at } \tau_f \sim 1, \\ 17.3 \text{ h} \left(\frac{\eta}{0.04}\right)^{-2} \left(\frac{\rho_g}{40 \text{ kg m}^{-3}}\right)^{-1} \left(\frac{\rho_p}{3000 \text{ kg m}^{-3}}\right) \left(\frac{M}{M_\oplus}\right)^{-\frac{1}{2}} \left(\frac{r}{3R_\oplus}\right)^{-\frac{1}{2}} \left(\frac{R_p}{2 \text{ km}}\right) & \text{at } \tau_f \gg 1. \end{cases} \quad (2)$$

$G$ is the gravitational constant, and $M$ is the mass of the planet. This leads to $t_{\text{fall}} = 1.21$ days at $R_p = 2$ km ($\tau_f = 1.18$), $t_{\text{fall}} = 72$ days at $R_p = 200$ km ($\tau_f = 118$), $t_{\text{fall}} \sim 1.68$ years at $R_p = 1700$ km ($\tau_f \sim 1010$) (the radius of the Moon is $1.731 \times 10^3$ km), and $t_{\text{fall}} \sim 280$ years at $R_p \sim 1$ cm ($\tau_f = 5.9 \times 10^{-6}$ and $r = 2R_\oplus$ is assumed). This indicates that any moonlets that reach ~km in size will fall inwards on a very short timescale (less than 1 day). For an icy moon-forming disk, $\tau_f = 1$ at $R_p = 1.3$ km assuming $\rho_p = 1000$ kg m$^{-3}$ and $\rho_g = 10$ kg m$^{-3}$. We assume that liquid droplets can accrete and grow quickly and that turbulence at the midplane would not prevent growth[36].

During the radial fall, a moonlet may evaporate due to heating by the gas drag. The total heating during the fall is approximately written as $Fv_{\text{vel}} t_{\text{fall}}$. The evaporative mass fraction of the moonlet is written as $Fv_{\text{vel}} t_{\text{fall}}/M_{\text{moonlet}} L$, where $M_{\text{moonlet}}$ is the moonlet mass ($=\frac{3}{4} \rho_p \pi R_p^3$) and $L$ is the latent heat. Evaporation is not a significant factor for a 2 km rocky moonlet in the moon-forming disk; only 20 wt% of the moonlet evaporate, whereas the icy moonlet completely evaporates

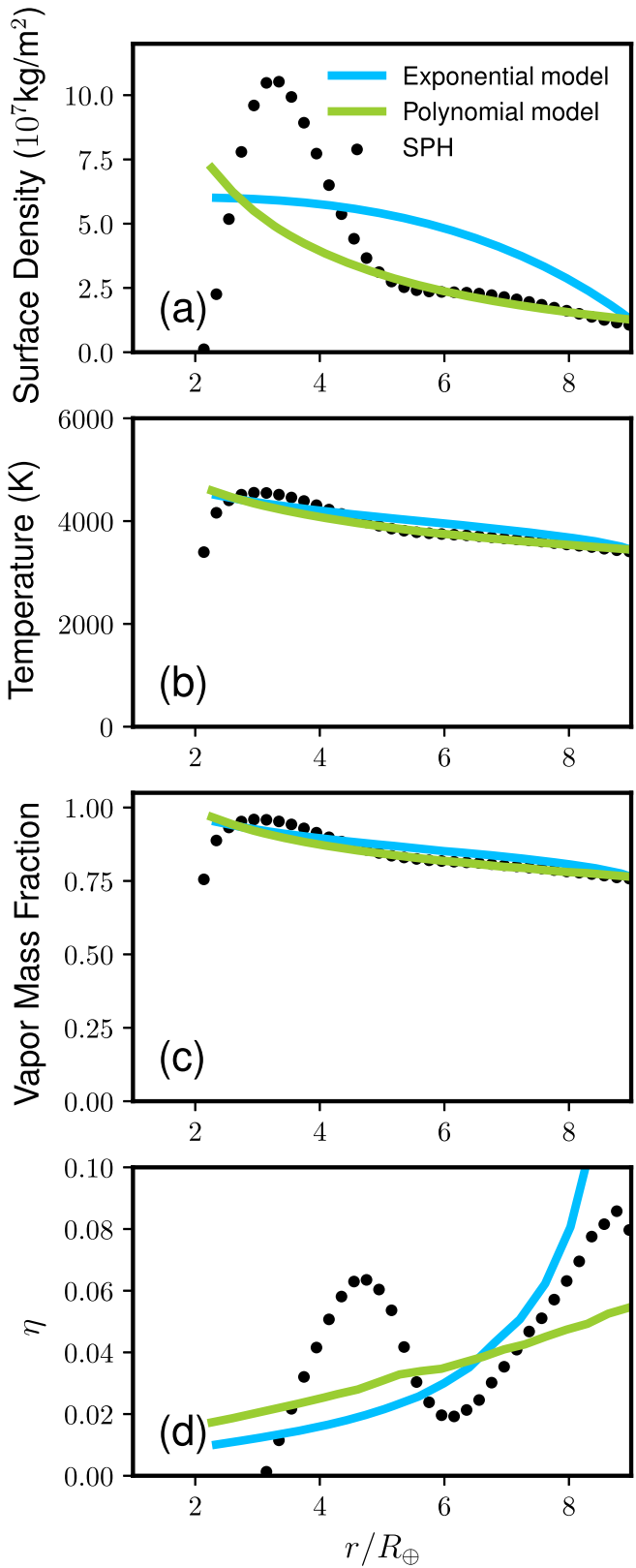

**Fig. 4 Disk structure formed by an impact between rocky planets** ($M_T = 5M_\oplus$, $\gamma = 0.13$, $\theta = 48.5°$, $v_{imp} = v_{esc}$, and Run ID6). All of these models have the same angular momenta and disk masses. The panels show (**a**) the surface density, (**b**) temperature, (**c**) vapor mass fraction, and (**d**) pressure gradient parameter $\eta$ as a function of the radial distance from the planet $r$ normalized by the Earth radius $R_\oplus$. The dotted line represents the direct outcome from the SPH simulation. The sky-blue line represents an exponential model and the green line represents a polynomial disk model. The planetary radius is 1.74 $R_\oplus$.

**Disk evolution.** Here, we describe a simple model that tracks the time evolution of a completely vapor disk. The disk would viscously spread while conserving the angular momentum. The analytical solution of the surface density of the disk is described as[49]

$$\Sigma = \Sigma_0 t_{*0}^{-21/22} \left(\frac{r}{R}\right)^{-3/4} \exp\left[-\left(\frac{r}{r_{d0}}\right)^{5/4} t_{*0}^{-15/22}\right], \quad (3)$$

where

$$t_{*0} = 1 + \frac{t}{t_{diff0}} = 1 + \frac{t}{\frac{16}{75}\left(\frac{r^2}{\nu}\right)_{r_{d0}, t=0}}. \quad (4)$$

$\Sigma_0$ is the initial surface density at $r = R$, where $r$ is the orbital distance and $R$ is the planetary radius. $r_{d0}$ is the typical disk size, which we define 3$R$. $t$ is time, and $t_{diff0}$ is the diffusion time at $r = r_{d0}$ and $t = 0$ and $\nu$ is the viscosity. Traditionally, the viscosity of the moon-forming disk is assumed to be controlled either by the instability-driven viscosity[61,62] or thermally driven viscosity[44]. The instability-driven viscosity is caused by self-gravitating clumps forming from a gravitationally unstable liquid layer inside the Roche radius. Inside the Roche radius, any self-gravitating bodies will be destroyed by the planetary tide. The thermally driven viscosity is caused by balancing heat generation by gravitational instability and radiative cooling. Here, we consider a completely vapor disk, which is gravitationally stable, and therefore no clumping is possible. Thus, the viscosities described above are not representative in our scenario. Here, we assume the turbulent kinematic viscosity described as $\nu = \alpha c_s^2 \Omega_K^{-1}$ (see refs. [63,64]), where $\alpha$ is a parameter that describes the turbulent strength, $c_s$ is the sound speed, and $\Omega_K$ is the Keplerian angular velocity. $\alpha$ is unknown, but $10^{-3}$ has been used for a protoplanetary disk and $10^{-6}-10^{-2}$ has been used for the magneto-rotational instability (MRI)-driven lunar forming disk[65–67]. Here, we use $\alpha = 5 \times 10^{-3}$ as a conservative value. $c_s = \sqrt{\gamma R_{gas} T/m}$, where $\gamma$ is the heat capacity ratio ($\tilde{\gamma} = 1.4$), $R_{gas}$ is the gas constant, $T$ (= 4000 K for rocky planets and 1000 K for icy planets) is the temperature, $m$ is the mean molecular weight (30 g/mol for silicate[44] and 18 g/mol for water).

The timescale for the disk lifetime depends on the model. For the Moon-forming disk, it is often considered to be ~100 years[43,44] but the timescale estimates range from 10s of years to 1000s of years[16,44,64–66]. Here, we assume that the disk lifetime is comparable to the disk condensation timescale. Roughly speaking, the timescale $\tau$ is described as

$$\tau = \frac{fML}{16\pi R^2 \sigma_{SB} T_{ph}^4} = 100 \left(\frac{f}{0.02}\right)\left(\frac{M}{M_\oplus}\right)\left(\frac{R}{R_\oplus}\right)^{-2}\left(\frac{L}{1.2 \times 10^7 \text{ J kg}^{-1}}\right)\left(\frac{T_{ph}}{1410 \text{ K}}\right)^{-4} \text{years}, \quad (5)$$

where $\sigma_{SB}$ is the Stefan Boltzmann constant, $f$ is the disk mass with respect to the planetary mass $M$, and $R$ is the planetary radius, where it is assumed that the typical disk radius is 3$R$. We also assume $M \propto R^{3.7}$ for rocky planets[28] and $M \propto R^3$ for icy planets[47]. $L$ is the latent heat, where $L = 2.3 \times 10^6$ J kg$^{-1}$ for icy planets and $L = 1.2 \times 10^7$ J kg$^{-1}$ for rocky planets. $T_{ph}$ is the photospheric temperature. Traditionally, the photospheric temperature of the Moon-forming disk has been assumed to be ~2000 K[44], but the temperature can be lower if heat transport to the photosphere is not efficient[65]. In this calculation, we assume the photospheric temperature for silicate is 1410 K, which gives $\tau \sim 100$ years for rocky systems. $T_{ph}$ is 270 K for icy planets[68].

Figure 7a shows the surface densities of the Moon-forming disk at 0 year (purple solid line), 10 years (light purple dash-dot line), and 100 years (yellow dashed line), assuming $f = 0.02$, $M = M_\oplus$, $R = R_\oplus$, $L = 1.2 \times 10^7$ J kg$^{-1}$, $T_{ph} = 1410$ K. The $x$-axis represents the distance from the planet normalized by the Earth radius $R_\oplus$. As time passes, the disk extends radially and the surface density decreases. After 100 years, the disk mass becomes 0.155 of the original disk mass. Figure 7b shows the final disk mass $M_{D,final}$ normalized by the initial disk mass $M_{D,initial} = fM$ as a function of various planetary masses ($f = 0.02$). The purple solid line represents icy planets, and the yellow dashed-dotted line represents rocky planets. The disk evolution calculations are stopped when the disk lifetime is reached. $M_{D,final}/M_{D,initial} \sim 0.2$ for rocky planets in $M = 0.1-6~M_\oplus$, whereas $M_{D,final}/M_{D,initial} < 0.06$ for icy planets in the same mass range. This is because the lifetime of an icy moon-forming disk is longer than that of a rocky moon-forming

($\rho_p = 1000$ kg m$^{-3}$, $\rho_g = 10$ kg m$^{-3}$, $R_p = 1.3$ km). In this case, moonlets that evaporate increase the vapor fraction of the disk, slowing down the cooling process. It is possible that even more mass is lost to the planet during the cooling process. In addition, small droplets (mm and cm in size) falling from the outer (colder) part of the disk evaporate quickly due to the temperature difference between the droplets and vapor[16], but this evaporation process is slow and therefore not significant for km-sized moonlets.

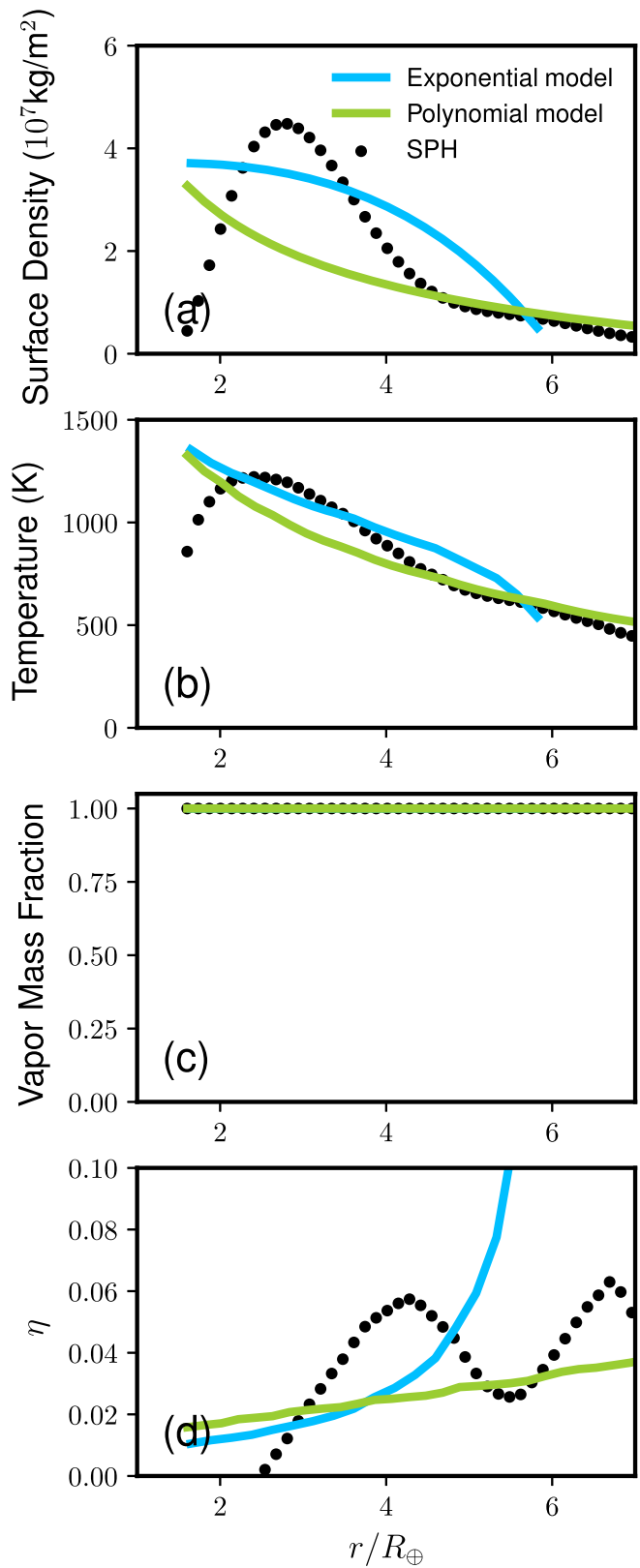

**Fig. 5 Disk structure at the midplane formed by an impact between icy planets ($M_T = 1 M_⊕$, $\gamma = 0.13$, $\theta = 48.5°$, $v_{imp} = v_{esc}$, Run ID20).** The line descriptions are the same as those in Fig. 4. The panels show (**a**) the surface density, (**b**) temperature, (**c**) vapor mass fraction, and (**d**) pressure gradient parameter $\eta$ as a function of $r/R_⊕$. The planetary radius is 1.42 $R_⊕$.

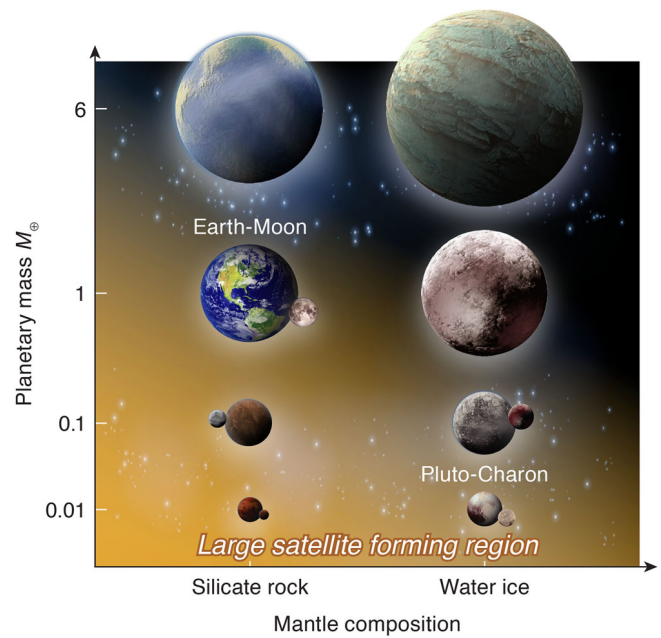

**Fig. 6 Schematic view of the mass range in which a fractionally large exomoon can form by an impact.** The horizontal axis represents the mantle composition and the vertical axis represents the planetary mass normalized by the Earth mass $M_⊕$. Rocky planets smaller than 6 $M_⊕$ and icy planets smaller than 1 $M_⊕$ are capable of forming fractionally large moons as indicated by the orange shading. Our prediction is consistent with planet–moon systems in the solar system.

disk due to its much smaller $T_{ph}$. Thus, the disk loses its significant portion of the mass by viscous evolution. Needless to say, this is a very simple model that ignores processes such as condensation, vapor-droplet interaction, and the Roche radius. More detailed investigations are needed to track the evolution of the disk.

**SPH simulations setting.** We use an SPH code that we developed from the ground up. This code has been extensively tested[18,20,35]. We use the semi-analytical equation of state called M-ANEOS to represent silicate rock (forsterite) and iron. The forsterite EOS we use here is "SPH-N" in previous work[69], where VMF can depend on the choice of an M-ANEOS input file. For ice, we used the five-phase EOS for water[70] for the range between 0 and 20,000 K and density smaller than 5000 kg m$^{-3}$. Outside of this range, the water EOS is interpolated using the table and SESAME 7154 EOS.

In most of the simulations, the initial entropy values of the mantles are 3165 J K$^{-1}$ kg$^{-1}$ and 3696 J K$^{-1}$ kg$^{-1}$ for rocky and icy planets, respectively. These values correspond to surface temperatures ~2000 K and 300 K for an Earth-sized planet, respectively. The exception is Run ID1 (Table 1) uses 1096 J K$^{-1}$ kg$^{-1}$. This corresponds to ~300 K at the surface.

The vapor mass fraction of a disk is determined assuming that the disk has a uniform entropy regardless of the radial distance and that the disk is in a hydrostatic equilibrium. The detailed method is described in our previous publication[20]. The disk's mass, surface density, and angular momentum are directly calculated from the SPH simulation and are listed in Table 1.

**Disk model.** Figure 4 shows the disk structure for a rocky planetary impact. The skyblue line represents an exponential surface density model ($\sigma(r) = (c_1 + c_2 r)\exp(-c_3 r)$) and here we pick values that satisfy $d\sigma/dr = 0$ at the inner edge, where $r$ is the radial distance from the Earth's spin axis. We conserve the disk mass and angular momentum within 5% ($c_1 = 5.857 \times 10^7$ kg m$^{-2}$, $c_2 = -0.948$ m$^{-1}$, and $c_3 = -2.023 \times 10^{-8}$ m$^{-1}$). The green line represents a polynomial model ($\sigma(r) = c_1 (r/r_0)^{c_2} + c_3$) and the error is less than 2% ($c_1 = 1.0 \times 10^8$ kg m$^{-2}$, $c_2 = -0.809$, $c_3 = -1.617 \times 10^7$ kg m$^{-2}$). Similarly, the structures of an icy disk shown in Fig. 5. The parameters we use for the exponential disk are $c_1 = 2.6505 \times 10^7$ kg m$^{-2}$, $c_2 = -0.5916$ m$^{-1}$, $c_3 = -2.79017 \times 10^{-8}$ m$^{-1}$ and those for the polynomial model are $c_1 = 5.5599 \times 10^7$ kg m$^{-2}$, $c_2 = -0.5060$, $c_3 = 1.920 \times 10^7$ kg m$^{-2}$. The error is less than 5% and 2%, respectively.

**$M_T$ dependence of VMF and local clumps.** Roughly speaking, the vapor mass fraction produced by an impact is scaled as $\frac{\gamma}{(1-\gamma)L} v_{imp}^2 \propto \frac{\gamma}{(1-\gamma)L} M_t^{2/3}$, where $M_t$ is the target mass, $L$ is the latent heat, and $v_{imp}$ is the impact velocity, which is assumed to be proportional to the escape velocity of the target, $\sqrt{2GM_t/R_t}$. For simplicity, $R_t \propto M_t^{1/3}$ is assumed and $\gamma$ is assumed to be small. Therefore, larger planets and more ice-rich planets produce high VMF disks in general.

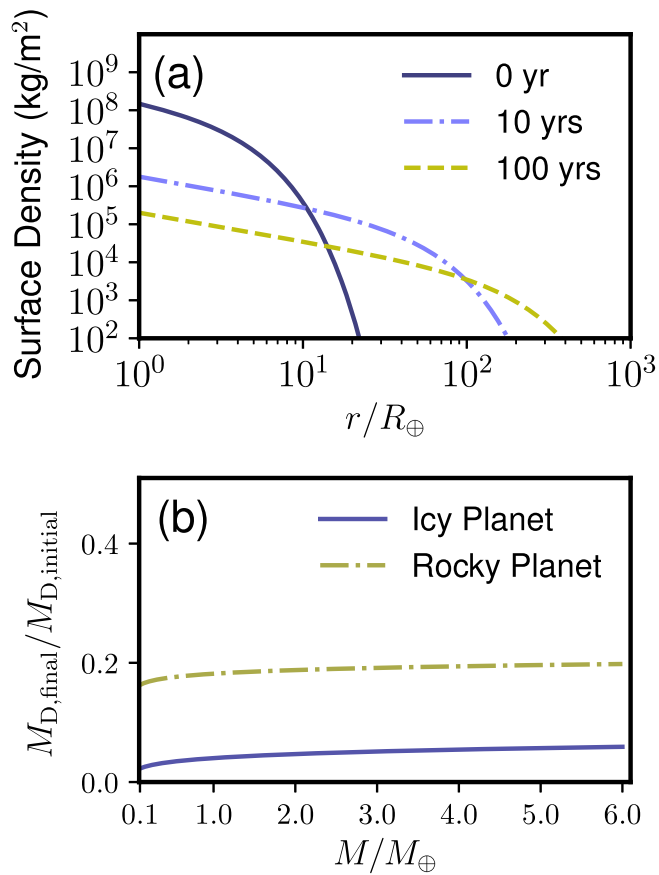

**Fig. 7 Evolution of completely vapor disks. a** Evolution of the surface density of the Moon-forming disk as a function of $r/R_\oplus$. The purple solid line, light purple dashed-dotted line, and the yellow dashed line represent the surface density at 0 year, 10 years, and 100 years, respectively. **b** The final disk mass normalized by the initial disk mass for icy planets (solid purple line) and rocky planets (dashed yellow line) as a function of the planetary mass M normalized by $M_\oplus$. For the details, see the main text.

It is possible to have the non-negligible amount of solid or liquid in the disk if its VMF is smaller than 1 and $M_T$ is large. For example, in Run ID19, the disk liquid mass is $0.129 \times 6M_\oplus \times (1-\gamma) \times (1-\text{VMF}) \sim 0.001\,M_\oplus$. If the entire mass is in one object, the radius is ~640 km. Its residence time would be ~230 days, which can be large enough to make a difference in the moon accretion process. Nevertheless, this is an unlikely scenario given that VMF tend to be uniform throughout the disk (Figs. 4 and 5 as well as previous work[20]), making it unlikely to have a local concentration of melt.

### Data availability
Data produced in this study are summarized in Table 1. Raw data are not shown due to their large size but are available from the corresponding author upon reasonable request.

### Code availability
The numerical codes that are used in this study are available from the corresponding author upon reasonable request.

## Acknowledgements
We thank David J. Stevenson, John Tarduno, David Kipping, and Alycia Weinberger for the insightful discussion. M.N. was supported in part by the National Aeronautics and Space Administration (NASA) Headquarters under the NASA Earth and Space Science Fellowship Program Grant NNX14AP26H, the Carnegie DTM fellowship, NASA grant numbers 80NSSC19K0514 and 80NSSC21K1184 and Grant-in-Aid for JSPS Fellows JSPS KAKENHI Grant Number 10J09549. Partial funding for M.N. was provided by the Center for Matter at Atomic Pressures (CMAP), a National Science Foundation (NSF) Physics Frontier Center, under Award PHY-2020249. Any opinions, findings, conclusions or recommendations expressed in this material are those of the authors and do not necessarily reflect those of the National Science Foundation. M.N. was also supported in part by the Alfred P. Sloan Foundation under grant number G202114194. S.I. was supported by JSPS Kakenhi 21H04512. H.G. was supported by MEXT KAKENHI Grant No. JP17H06457. Numerical computations were partly carried out on GRAPE system at Center for Computational Astrophysics, the National Astronomical Observatory of Japan. Figure 6 is created by Michael Franchot.

## Author contributions
M.N. contributed to the idea of the project, performed the simulations, and wrote the manuscript. H.G. contributed to the project design and the SPH code development, E.A. contributed to the EOS implementation, and S.I. contributed to the disk dynamics and evolution discussion. All authors contributed to the discussion, interpretation of the results, and revising the manuscript.

## Competing interests
The authors have no competing interests.

## Additional information
**Correspondence** and requests for materials should be addressed to Miki Nakajima.

**Peer review information** *Nature Communications* thanks the anonymous reviewer(s) for their contribution to the peer review of this work.

**Reprints and permission information** is available at http://www.nature.com/reprints

**Publisher's note** Springer Nature remains neutral with regard to jurisdictional claims in published maps and institutional affiliations.